\documentclass[twoside,twocolumn,9pt]{article}
\usepackage{extsizes}
\usepackage[super,sort&compress,comma]{natbib} 
\usepackage[version=3]{mhchem}
\usepackage[left=1.5cm, right=1.5cm, top=1.785cm, bottom=2.0cm]{geometry}
\usepackage{balance}
\usepackage{mathptmx}
\usepackage{sectsty}
\usepackage{graphicx} 
\usepackage{lastpage}
\usepackage[format=plain,justification=justified,singlelinecheck=false,font={stretch=1.125,small,sf},labelfont=bf,labelsep=space]{caption}
\usepackage{float}
\usepackage{fancyhdr}
\usepackage{fnpos}
\usepackage[english]{babel}
\addto{\captionsenglish}{%
  
}
\usepackage{array}
\usepackage{droidsans}
\usepackage{charter}
\usepackage[T1]{fontenc}
\usepackage[usenames,dvipsnames]{xcolor}
\usepackage{setspace}
\usepackage[compact]{titlesec}
\usepackage{hyperref}

\usepackage{epstopdf}

\usepackage{xfrac}
\usepackage{mhchem}
\usepackage{tabularx}

\definecolor{cream}{RGB}{222,217,201}

\begin{document}

\pagestyle{fancy}
\thispagestyle{plain}
\fancypagestyle{plain}{
\renewcommand{\headrulewidth}{0pt}
}

\makeFNbottom
\makeatletter
\renewcommand\LARGE{\@setfontsize\LARGE{15pt}{17}}
\renewcommand\Large{\@setfontsize\Large{12pt}{14}}
\renewcommand\large{\@setfontsize\large{10pt}{12}}
\renewcommand\footnotesize{\@setfontsize\footnotesize{7pt}{10}}
\makeatother

\renewcommand{\thefootnote}{\fnsymbol{footnote}}
\renewcommand\footnoterule{\vspace*{1pt}%
\color{cream}\hrule width 3.5in height 0.4pt \color{black}\vspace*{5pt}} 
\setcounter{secnumdepth}{5}

\makeatletter 
\renewcommand\@biblabel[1]{#1}            
\renewcommand\@makefntext[1]%
{\noindent\makebox[0pt][r]{\@thefnmark\,}#1}
\makeatother 
\renewcommand{\figurename}{\small{Fig.}~}
\sectionfont{\sffamily\Large}
\subsectionfont{\normalsize}
\subsubsectionfont{\bf}
\setstretch{1.125} 
\setlength{\skip\footins}{0.8cm}
\setlength{\footnotesep}{0.25cm}
\setlength{\jot}{10pt}
\titlespacing*{\section}{0pt}{4pt}{4pt}
\titlespacing*{\subsection}{0pt}{15pt}{1pt}

\fancyfoot{}
\fancyfoot[LO,RE]{\vspace{-7.1pt}\includegraphics[height=9pt]{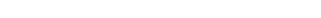}}
\fancyfoot[CO]{\vspace{-7.1pt}\hspace{13.2cm}\includegraphics{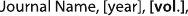}}
\fancyfoot[CE]{\vspace{-7.2pt}\hspace{-14.2cm}\includegraphics{head_foot/RF}}
\fancyfoot[RO]{\footnotesize{\sffamily{1--\pageref{LastPage} ~\textbar  \hspace{2pt}\thepage}}}
\fancyfoot[LE]{\footnotesize{\sffamily{\thepage~\textbar\hspace{3.45cm} 1--\pageref{LastPage}}}}
\fancyhead{}
\renewcommand{\headrulewidth}{0pt} 
\renewcommand{\footrulewidth}{0pt}
\setlength{\arrayrulewidth}{1pt}
\setlength{\columnsep}{6.5mm}
\setlength\bibsep{1pt}

\makeatletter 
\newlength{\figrulesep} 
\setlength{\figrulesep}{0.5\textfloatsep} 

\newcommand{\topfigrule}{\vspace*{-1pt}%
\noindent{\color{cream}\rule[-\figrulesep]{\columnwidth}{1.5pt}} }

\newcommand{\botfigrule}{\vspace*{-2pt}%
\noindent{\color{cream}\rule[\figrulesep]{\columnwidth}{1.5pt}} }

\newcommand{\dblfigrule}{\vspace*{-1pt}%
\noindent{\color{cream}\rule[-\figrulesep]{\textwidth}{1.5pt}} }

\makeatother

\twocolumn[
  \begin{@twocolumnfalse}
\vspace{1em}
\sffamily
\begin{tabular}{m{4.5cm} p{13.5cm} }

\includegraphics{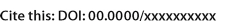} & \noindent\LARGE{\textbf{Varying A-site radius and size disorder to tune magnetic ordering in compositionally complex perovskite oxides$^\dag$}} \\
\vspace{0.3cm} & \vspace{0.3cm} \\

 & \noindent\large{Madeleine Geers,\textit{$^{a}$} Ravi Kiran Dokala,\textit{$^{b,c}$} Roland Mathieu\textit{$^{b}$} and Rebecca Clulow$^{\ast}$\textit{$^{a}$}} \\

\includegraphics{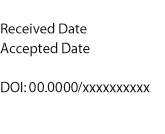} & \noindent\normalsize{A long standing goal of magnetochemists is to be able to control ordering temperatures through compositional engineering. Here we report the synthesis of a perovskite oxide with twelve cations accommodated within a single phase solid solution, AM7O$_3$, A = La$_{\sfrac{1}{5}}$Sm$_{\sfrac{1}{5}}$Gd$_{\sfrac{1}{5}}$Nd$_{\sfrac{1}{5}}$Dy$_{\sfrac{1}{5}}$ (A5), and M7 = Ti$_{\sfrac{1}{7}}$Cr$_{\sfrac{1}{7}}$Mn$_{\sfrac{1}{7}}$Fe$_{\sfrac{1}{7}}$Co$_{\sfrac{1}{7}}$Ni$_{\sfrac{1}{7}}$Cu$_{\sfrac{1}{7}}$. Bulk magnetometry measurements for A5M7O$_3$, as well as for A = La, Gd and La$_{\sfrac{1}{2}}$Gd$_{\sfrac{1}{2}}$, show that all these compounds magnetically order as ferrimagnets between 89 and 115 K. We find that the magnetic properties are influenced by considering a combination of the size disorder parameter alongside the A-site cationic radii and valence electrons. } \\

\end{tabular}

 \end{@twocolumnfalse} \vspace{0.6cm}

  ]

\renewcommand*\rmdefault{bch}\normalfont\upshape
\rmfamily
\section*{}
\vspace{-1cm}


\footnotetext{\textit{$^{a}$~Department of Chemistry - Ångström laboratory, Uppsala University, Box 538, 751 21, Uppsala, Sweden; E-mail: rebecca.clulow@kemi.uu.se}}
\footnotetext{\textit{$^{b}$~Department of Materials Science and Engineering, Uppsala University, Box 523, 751 20, Uppsala, Sweden }} 
\footnotetext{\textit{$^{c}$~Department of Physics, Stockholm University, 106 91 Stockholm, Sweden }} 





\section{Introduction}

\begin{figure}[t]
\centering
  \includegraphics{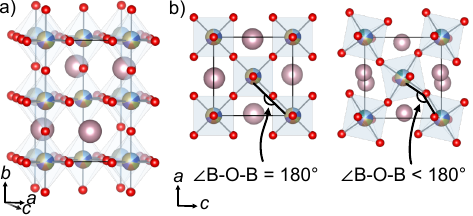}
  \caption{Structure of an undistorted perovskite. ABO$_3$, A = pink, B = multicoloured, O = red. b) The B-O-B angle in an undistorted perovskite (left, \textit{t} = 1), and one with a smaller A-site cation (right, \textit{t} < 1), where rotation of the octahedra and displacements of the A-site usually occurs.}
  \label{fgr:intro}
\end{figure}

The physical properties of many materials are inherently engendered by their crystal structure. Therefore, tailoring a material for specific functions should begin with investigations of the chemical composition and structural relationships. Perovskite oxides, with ABO$_3$ stoichiometry, are ubiquitous with promising magnetic, electrical, and optical properties.\cite{zhang_magnetic_2022, khosrozadeh_complex_2024, belguenoune_structural_2026} Built from corner-sharing [BO$_6$] octahedra and A-site cations positioned in the voids (Figure \ref{fgr:intro}a), this structure type offers a robust and versatile scaffold to explore compositional space.

From the archetypical cubic perovskite, one method to introduce structural distortions is through the comparative ratio of the A- and B-site radii, commonly expressed with Goldschmidt’s tolerance factor (\textit{t}).\cite{Goldschmidt1926} Compared to an ideal, undistorted perovskite with \textit{t}=1, perovskites incorporating smaller A-site cations, where \textit{t}<1, tend to induce rotations of the [BO$_6$] octahedra. Consequently, they typically adopt orthorhombic symmetry, as observed for the majority of rare earth-based perovskites. From a magnetic perspective, this can be advantageous, as, for example, the canting angle of weak ferromagnets has been demonstrated to be controlled by the extent of the octahedral tilts.\cite{zhao_effect_2013}

For rare-earth transition metal based perovskites, it is well established that the magnetic ordering temperature is predominately dictated by the transition metal B-sites, which are governed by the cations, orbital overlap and B$-$O$-$B exchange pathways.\cite{zhou_intrinsic_2010, terakura_magnetism_2007} Structural distortions which alter the B$-$O$-$B geometry arise through cooperative octahedral rotations, deviating from 180$^\circ$ in an undistorted structure, and are coupled with a lowering of the ordering temperature (Figure \ref{fgr:intro}b).\cite{lyubutin_dependence_1999, treves_dependence_1965, zhou_intrinsic_2008} Where Jahn-Teller active cations are present, a change in the B$-$O$-$B angle can alter the magnetic structure. This is axiomatic, for example, in AMnO$_3$ as a function of Mn$^{3+}-$O$-$Mn$^{3+}$ bond angle, where within a range of 4$^\circ$ the magnetic structure can adopt an incommensurate, A-type, or E-type antiferromagnetic structure.\cite{goto_ferroelectricity_2004, kimura_magnetoelectric_2005} 

Tuning through chemical substitution can be progressed to include, for example, upward of five constituent species within a single solid solution phase. These compositionally complex compounds allow for fine tuning of behaviour, both of their structural characteristics, such as control of phase transition temperatures,\cite{cedervall_phase_2021} and of their physical properties, including magnetism.\cite{witte_high-entropy_2019, das_comprehensive_2023} Moreover, it is proposed that the inclusion of several species with varied radii leads to structural distortions, which may positively influence ionic conduction\cite{Rajendran_tri-doped_2020} and mechanical properties.\cite{hong_microstructural_2019}

As a result of the high chemical disorder, compositionally complex materials provide a template whereby a relatively simplistic crystal structure can host a multitude of competing exchange interactions. This has ignited research on these materials as potential hosts for complex spin textures and exotic magnetic ground states. Despite the numerous magnetic interactions present within the single phase, these compounds usually adopt magnetically ordered ground states,\cite{ cedervall_phase_2021, witte_high-entropy_2019, das_comparative_2021, witte_magnetic_2020} with only limited examples of spin glasses being reported.\cite{Pramanik_spin_2024, clulow_phase_2024, arndt_magnetic_2024} 

An additional factor, when considering magnetic properties of compositionally complex materials, is the change associated with the range of cationic radii incorporated. Quantified by the size disorder parameter, $\sigma ^2 = \Sigma_i x_i r_i^2 - (\Sigma_i  x_i r_i)^2 = < r_\mathrm{A} ^2> - <r_\mathrm{A}> ^2$,\cite{rodriguez-martinez_structural_1999} where $x_i$ is the fraction of the \textit{i}th cation of radius $r_i$, it has been demonstrated that for compounds with a constant $<r_\mathrm{A}>$ (average ionic radius of the A-site), the magnetic transition temperature decreases linearly with increasing $\sigma ^2$.\cite{rodriguez-martinez_structural_1999, maignan_size_1999} Mapping of manganite perovskites demonstrate that the magnetic transitions and quenched disorder are influenced by both $\sigma ^2$ and $<r_\mathrm{A}>$,\cite{tomioka_global_2004} with suggestions that it likely has a larger influence on the magnetic ordering than solely $<r_\mathrm{A}>$.\cite{das_comprehensive_2023}

The majority of ABO$_3$ compounds with compositionally complex B-sites show weak ferromagnetic behaviour, with the ordering temperature correlating with $<r_\mathrm{A}>$ for a given B-site radius.\cite{cedervall_phase_2021, witte_high-entropy_2019, arndt_magnetic_2024, witte_magnetic_2020} Many of these compounds exhibit magnetic order as a gradual transition, which is assumed to arise from the large variations in the strength and sign of the competing B$-$O$-$B exchange interactions.\cite{witte_high-entropy_2019} Notably, LaM5O$_3$, M5 = Cr$_{\sfrac{1}{5}}$Mn$_{\sfrac{1}{5}}$Fe$_{\sfrac{1}{5}}$Co$_{\sfrac{1}{5}}$Ni$_{\sfrac{1}{5}}$, exhibits a vertical exchange bias with a significant remanent magnetisation and SmM7O$_3$, M7 = Ti$_{\sfrac{1}{7}}$Cr$_{\sfrac{1}{7}}$Mn$_{\sfrac{1}{7}}$Fe$_{\sfrac{1}{7}}$Co$_{\sfrac{1}{7}}$Ni$_{\sfrac{1}{7}}$Cu$_{\sfrac{1}{7}}$, demonstrates switchable magnetic states.\cite{witte_high-entropy_2019, dokala_switching_2026} Both of these compounds have their behaviour attributed to the intrinsic complexity introduced by their multi-cationic B-site. Despite the enticing magnetic results, only La$_{\sfrac{1}{2}}$Nd$_{\sfrac{1}{2}}$M7O$_3$ ($\sigma ^2 =$0.002) has thus far also been studied from the AM7O$_3$ family.\cite{cedervall_phase_2021} Given that almost all of the first row rare earth cations have been incorporated into transition metal perovskites, there is a vast chemical space left to explore when considering the magnetic tunability through $<r_\mathrm{A}>$ and $\sigma ^2$.

In this work, we present four compositionally complex perovskite oxides, AM7O$_3$, A = La, Gd, La$_{\sfrac{1}{2}}$Gd$_{\sfrac{1}{2}}$ and  La$_{\sfrac{1}{5}}$Sm$_{\sfrac{1}{5}}$Gd$_{\sfrac{1}{5}}$Nd$_{\sfrac{1}{5}}$Dy$_{\sfrac{1}{5}}$ (A5), and M7 = Ti$_{\sfrac{1}{7}}$Cr$_{\sfrac{1}{7}}$Mn$_{\sfrac{1}{7}}$Fe$_{\sfrac{1}{7}}$Co$_{\sfrac{1}{7}}$Ni$_{\sfrac{1}{7}}$Cu$_{\sfrac{1}{7}}$. We describe the synthesis of A5M7O$_3$, a single phase perovskite incorporating a total of twelve A- and B-site species. The structure and composition were determined with powder X-ray diffraction and energy dispersive X-ray spectroscopy (EDS) data. We used bulk magnetometry measurements to show that all four of these compounds magnetically order, with the magnetic properties closely coupled to the A-site species incorporated. Our results suggest that compositionally complex materials offer a promising foundation to explore fine-control of the physical properties of materials through compositional tuning.

\section{Experimental}

\subsection{Synthesis}

Samples were prepared using traditional solid-state synthesis methods. Lanthanide oxides (\ce{La2O3}, \ce{Nd2O3}, \ce{Sm2O3}, \ce{Gd2O3}, \ce{Dy2O3}) with purity $\geq{99.9 \%} $ were pre-dried at $975^\circ$C prior to weighing and mixed in stoichiometric proportions with the transition metal oxides (\ce{TiO2}, \ce{Cr2O3}, MnO, \ce{Fe2O3}, \ce{Co3O4}, NiO, CuO) with purity $\geq{99.5 \%}$. The mixtures were ground in minimal ethanol (approx. $1$ mL) and pressed into pellets before heating. The pellets were heated in air at $900 ^\circ$C for $12$h, $1000 ^\circ$C for $24$h, and $1100 ^\circ$C for $48$h. An additional heating step was carried out for LaM7O$_3$ at $1180 ^\circ$C for $48$h. During sintering, a heating rate of $5 ^\circ$C/min and a cooling rate of $10 ^\circ$C/min were used.

\subsection{Powder X-ray diffraction}

Powder X-ray diffraction (PXRD) patterns were collected at room temperature using a Bruker D8 ADVANCE diffractometer equipped with TWIN/TWIN setup (40 kV, 40 mA) using Cu K$\alpha$ radiation in Bragg-Brentano geometry. Data were collected using a Lynx-eye XE-T position sensitive detector, operated over an angular 2$\theta$ range of 10$-$80$^\circ$ with a step size of 0.02$^\circ$. Rietveld refinements of the structural models were carried out using the TOPAS software suite.\cite{coelho_topas_2018} Unit cell parameters, background and peak shape parameters were freely refined. Atomic coordinates for the B-site metals (M7 = Ti, Cr, Mn, Fe, Co, Ni, Cu), and for the A-site where A = La$_{\sfrac{1}{2}}$Gd$_{\sfrac{1}{2}}$ and A5, were constrained so that the atomic coordinates were equivalent. Displacement parameters were fixed throughout the refinements. 

\subsection{EDS}

The sample compositions and morphology were obtained using a ZEISS Leo 1550 field emission scanning electron microscope (SEM) equipped with an AZtec energy dispersive X-ray detector for spectroscopy analysis (EDS) on 10 spots using an accelerating voltage of 20 kV. Pellets of the sintered powders were attached to conducting carbon tape and point analysis and mapping were performed.

\subsection{Magnetic measurements}

The temperature-dependent zero field-cooled (ZFC), field-cooled (FC) magnetisation was collected using a squid magnetometer from Quantum Design Inc. Magnetic field (\textit{H}) dependent magnetisation (\textit{M}) measurements were recorded on the same equipment at a constant temperature after cooling the samples in ZFC and FC conditions.

\section{Results}
\subsection{Structure}

\begin{figure*}[t]
\centering
  \includegraphics{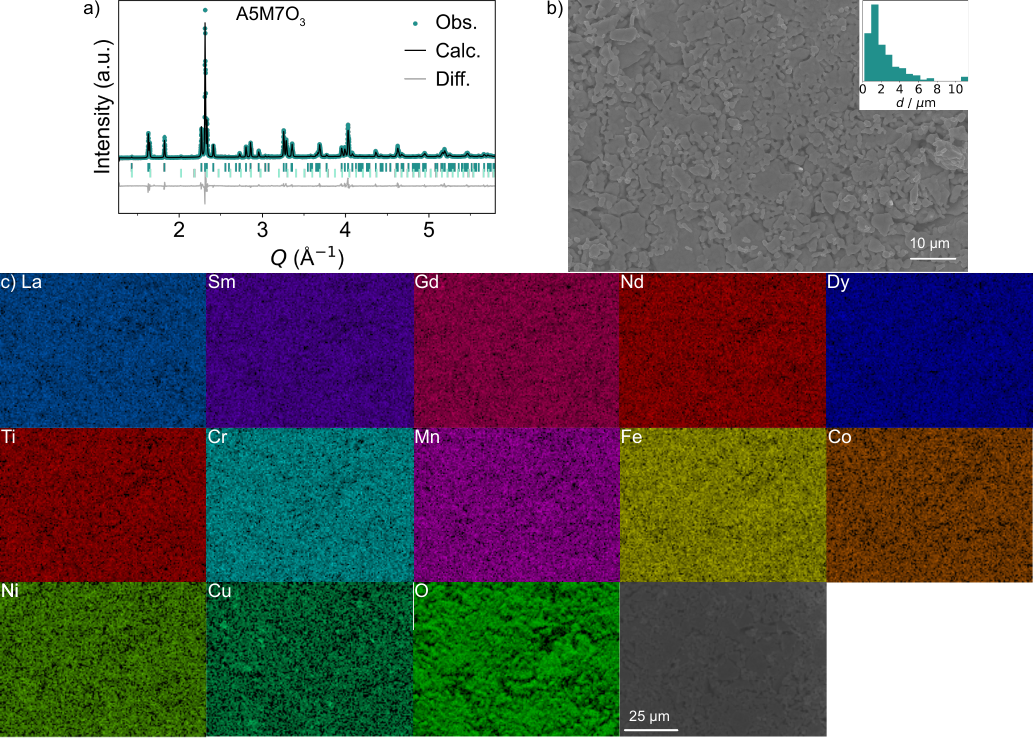}
  \caption{Structural characterisation of A5M7O$_3$. a) The Rietveld refinement results from powder X-ray diffraction data with $Pnma$ symmetry (dark teal tick marks) and \ce{Gd2O3} impurity phase (0.22 weight \%, light teal), $R_\mathrm{wp} =$ 6.7 \%. b) SEM image. c) Elemental distribution of the constituent elements from EDS mapping.}
  \label{fgr:structure}
\end{figure*}

\begin{table*}[h]
\small
  \caption{\ A-site radii ($r_\mathrm{A}$, XII coordination),\cite{baloch_extending_2021} calculated size disorder parameter, $\sigma ^2$, and refined unit cell parameters and statistics obtained from powder X-ray diffraction data for the compounds AM7O$_{3}$, M=Ti$_{\sfrac{1}{7}}$Cr$_{\sfrac{1}{7}}$Mn$_{\sfrac{1}{7}}$Fe$_{\sfrac{1}{7}}$Co$_{\sfrac{1}{7}}$Ni$_{\sfrac{1}{7}}$Cu$_{\sfrac{1}{7}}$; and A = La, Gd, La$_{\sfrac{1}{2}}$Gd$_{\sfrac{1}{2}}$, La$_{\sfrac{1}{5}}$Sm$_{\sfrac{1}{5}}$Gd$_{\sfrac{1}{5}}$Nd$_{\sfrac{1}{5}}$Dy$_{\sfrac{1}{5}}$ (A5). All compounds adopt \textit{Pnma} symmetry}
  \label{tbl:UC_param}
  \begin{tabularx}{\textwidth}{@{\extracolsep{\fill}}llllllll} 
    \hline
    A-site cation & $r_\mathrm{A}$ (Å) & $\sigma ^2$ & \textit{a} (Å) & \textit{b} (Å) & \textit{c} (Å) & \textit{V} (Å$^3$) & $R_\mathrm{wp}$ (\%)  \\
    \hline
    Gd  & 1.21 & 0 & 5.57971(7) & 7.5866(1) & 5.31611(7) & 225.037(5) & 6.17  \\
    La$_\frac{1}{2}$Gd$_\frac{1}{2}$ & 1.28  & 0.00563 & 5.53039(9) & 7.70419(13) & 5.43054(10) & 231.380(7) & 7.04  \\
    A5  & 1.25 & 0.00354 & 5.54152(10) & 7.64919(13) & 5.38185(10) & 228.127(7) & 7.30  \\
    La	& 1.36 & 0 & 5.49751(7) & 7.78107(9) & 5.52357(7) & 236.279(5)  & 7.61 \\
    \hline 
  \end{tabularx}
\end{table*} 

\begin{table*}[h]
\small
  \caption{\ Elemental compositions obtained from energy dispersive X-ray spectroscopy (EDS) measurements}
  \label{tbl:EDS_comp}
  \begin{tabularx}{\textwidth}{@{\extracolsep{\fill}}lll} 
    \hline
    A-site cation & Nominal composition & Average EDS composition \\
    \hline
    Gd & Gd$_{1.0}$Ti$_{0.142}$Cr$_{0.142}$Mn$_{0.142}$Fe$_{0.142}$Co$_{0.142}$Ni$_{0.142}$Cu$_{0.142}$ &  Gd$_{1.0(2)}$Ti$_{0.14(2)}$Cr$_{0.16(1)}$Mn$_{0.15(1)}$Fe$_{0.16(1)}$Co$_{0.15(2)}$Ni$_{0.142(8)}$Cu$_{0.101(9)}$ \\
    La$_\frac{1}{2}$Gd$_\frac{1}{2}$ & La$_{0.50}$Gd$_{0.50}$Ti$_{0.142}$Cr$_{0.142}$Mn$_{0.142}$Fe$_{0.142}$Co$_{0.142}$Ni$_{0.142}$Cu$_{0.142}$  &  La$_{0.52(2)}$Gd$_{0.47(2)}$Ti$_{0.15(2)}$Cr$_{0.18(3)}$Mn$_{0.11(2)}$Fe$_{0.15(2)}$Co$_{0.13(2)}$Ni$_{0.15(2)}$Cu$_{0.12(2)}$ \\
    A5 & La$_{0.20}$Sm$_{0.20}$Gd$_{0.20}$Nd$_{0.20}$Dy$_{0.20}$  &   La$_{0.19(1)}$Sm$_{0.20(1)}$Gd$_{0.19(1)}$Nd$_{0.21(1)}$Dy$_{0.20(2)}$ \\
     & Ti$_{0.142}$Cr$_{0.142}$Mn$_{0.142}$Fe$_{0.142}$Co$_{0.142}$Ni$_{0.142}$Cu$_{0.142}$ & Ti$_{0.14(3)}$Cr$_{0.14(1)}$Mn$_{0.17(4)}$Fe$_{0.15(2)}$Co$_{0.14(2)}$Ni$_{0.14(1)}$Cu$_{0.12(1)}$ \\
    La & La$_{1.0}$Ti$_{0.142}$Cr$_{0.142}$Mn$_{0.142}$Fe$_{0.142}$Co$_{0.142}$Ni$_{0.142}$Cu$_{0.142}$ &  La$_{1.0(1)}$Ti$_{0.14(2)}$Cr$_{0.14(2)}$Mn$_{0.14(2)}$Fe$_{0.15(1)}$Co$_{0.14(1)}$Ni$_{0.13(1)}$Cu$_{0.12(1)}$ \\
    \hline 
  \end{tabularx}
\end{table*}

The previously unreported A5M7O$_{3}$ compound, A5 = La$_{\sfrac{1}{5}}$Sm$_{\sfrac{1}{5}}$Gd$_{\sfrac{1}{5}}$Nd$_{\sfrac{1}{5}}$Dy$_{\sfrac{1}{5}}$, and M = Ti$_{\sfrac{1}{7}}$Cr$_{\sfrac{1}{7}}$Mn$_{\sfrac{1}{7}}$Fe$_{\sfrac{1}{7}}$Co$_{\sfrac{1}{7}}$Ni$_{\sfrac{1}{7}}$Cu$_{\sfrac{1}{7}}$, was synthesised by conventional solid state synthesis methods with a final heating temperature of 1100 $^{\circ}$C for 48 h. Synthesis of the subsequent compounds, A = La, Gd and La$_{\sfrac{1}{2}}$Gd$_{\sfrac{1}{2}}$, were carried out using the same heating conditions, similar to methods previously reported.\cite{cedervall_phase_2021} Rietveld refinements of powder X-ray diffraction for A5M7O$_{3}$ show that the compound is isostructural to the other AM7O$_3$ compounds in this series, crystallising in the orthorhombic \textit{Pnma} space group, with no evidence of additional A- or B-site orderings. Small impurities ($\le$0.3 \%) of \ce{Gd2O3} were present in GdM7O$_3$, La$_{\sfrac{1}{2}}$Gd$_{\sfrac{1}{2}}$M7O$_3$ and A5M7O$_3$. 

Generally, as the average A-site radius decreases, the unit cell volume decreases with unit cells of \textit{a} $\approx$ 5.5 Å, \textit{b} $\approx$ 7.6 Å and \textit{c} $\approx$ 5.4 Å (Table~\ref{tbl:UC_param}). There are two crystallographically distinct oxygen sites, which are axially (O1, Wyckoff site 4\textit{c}) and equitorially (O2, Wyckoff site 8\textit{d}) coordinated to the B-site cations. The \ce{[BO6]} octahedra are rotated relative to the principal axes, along the [$010$], [$10\Bar{1}$] and [$101$] directions, with a tilt sequence of \textit{a}$^-$\textit{a}$^-$\textit{c}$^+$ (Glazer notation\cite{glazer_classification_1972}), the most common sequence observed in orthorhombic perovskite structures. The magnitude of rotation of the \ce{[BO6]} units is correlated with the average A-site radius, with larger tilt angles observed where smaller A-site cations are incorporated, such as for GdM7O$_3$ with $\angle$B$-$O2$-$B = 149.6(4)\textdegree\ compared to 162.3(7)\textdegree\ for LaM7O$_3$ (Table S13).

To explore the efficacy of cation mixing and homogeneity of the powders, elemental analyses were carried out using energy dispersive X-ray spectroscopy (EDS) data. For all compounds, their compositions were close to their nominal values, with the B-site cation ratios ranging between 0.10 and 0.16 (expected = 0.142, Table \ref{tbl:EDS_comp}). For A5M7O$_3$, the A-site cations are homogeneously distributed, in quantities between 0.19(1) and 0.21(1), close to the expected value of 0.20 (Figure \ref{fgr:structure}c). In addition, corresponding microscopic images were collected from scanning electron microscopy (SEM), showing the compounds have an average particle size between 2 and 4 µm. The results suggest that all the compounds have a homogeneous distribution of both the A- and B-site cations on a micro-length scale.

\subsection{Magnetism}

\begin{table}[!ht]
    \centering
    \caption{Magnetic parameters obtained from magnetisation measurements for the compounds AM7O$_{3}$, M=Ti$_{\sfrac{1}{7}}$Cr$_{\sfrac{1}{7}}$Mn$_{\sfrac{1}{7}}$Fe$_{\sfrac{1}{7}}$Co$_{\sfrac{1}{7}}$Ni$_{\sfrac{1}{7}}$Cu$_{\sfrac{1}{7}}$; and A = La, Gd, La$_{\sfrac{1}{2}}$Gd$_{\sfrac{1}{2}}$, La$_{\sfrac{1}{5}}$Sm$_{\sfrac{1}{5}}$Gd$_{\sfrac{1}{5}}$Nd$_{\sfrac{1}{5}}$Dy$_{\sfrac{1}{5}}$ (A5). Magnetic ordering temperature ($T_\mathrm{N}$), Curie–Weiss temperature ($\theta_\mathrm{CW}$), experimentally determined effective magnetic moment ($\mu_\mathrm{eff}$), and the A-site effective moment ($\mu_\mathrm{eff, \; A}$), theoretically estimated from the rare-earth contribution}
    \begin{tabular}{lrrrc}
    \hline
        A-site cation & $T_\mathrm{N}$ (K) & $\theta_\mathrm{CW}$ (K) & $\mu_\mathrm{eff}$ ($\mu_\mathrm{B}$) & $\mu_\mathrm{eff, \; A}$ ($\mu_\mathrm{B}$) \\ \hline
        Gd & 89(2) & $-$9(2) & 8.9(1) & 7.94 \\
        La$_\frac{1}{2}$Gd$_\frac{1}{2}$ & 95(2) & +4(2) & 6.2(1) & 5.61 \\
        A5 & 90(2) & +14(2) & 6.2(1) & 6.16 \\ 
        La & 115(2) & +6(2) & 3.1(1) & 0.00 \\ \hline
    \end{tabular}
    \label{mag_values}
\end{table}

\begin{figure}[t]
\centering
  \includegraphics{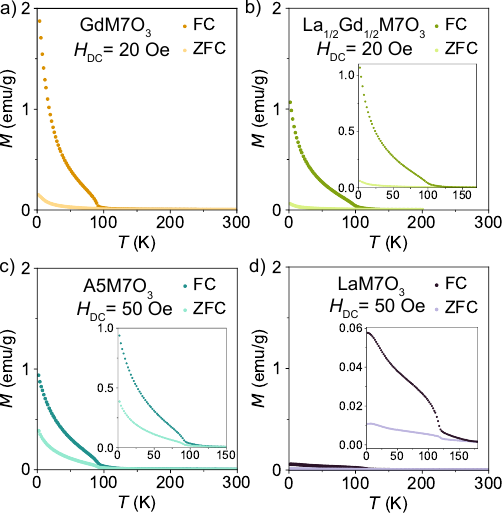}
  \caption{Temperature dependent magnetisation under zero-field-cooled (ZFC) and field-cooled (FC) protocols for a) GdM7O$_3$, b) La$_{\sfrac{1}{2}}$Gd$_{\sfrac{1}{2}}$M7O$_3$, c) A5M7O$_3$ and d) LaM7O$_3$. ZFC and FC measurements were performed under applied magnetic fields of 20 or 50 Oe.}
  \label{fgr:susceptibility}
\end{figure}

\begin{figure}[h]
\centering
  \includegraphics{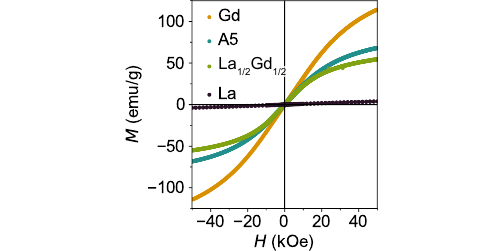}
  \caption{Isothermal magnetisation measurements carried out at 2 K for GdM7O$_3$ (orange), A5M7O$_3$ (teal),  La$_{\sfrac{1}{2}}$Gd$_{\sfrac{1}{2}}$M7O$_3$ (green), and LaM7O$_3$ (purple).}
  \label{fgr:magnetisation}
\end{figure}

To explore the effect of the A-site composition on their magnetic properties, bulk magnetometry measurements were carried out. Direct current (dc) magnetisation measurements were performed under zero-field-cooled (ZFC) warming and field-cooled (FC) cooling protocols for LaM7O$_3$, GdM7O$_3$, La$_{\sfrac{1}{2}}$Gd$_{\sfrac{1}{2}}$M7O$_3$ and A5M7O$_3$ (Figure \ref{fgr:susceptibility}). All compounds exhibit pronounced inflections in their ZFC/FC curves at approximately $T_\mathrm{N}=$ 90–120 K, accompanied by magnetic irreversibility (divergence between the ZFC and FC magnetisation curves), which is characteristic of long range ferrimagnetic ordering below these temperatures (Table \ref{mag_values}).\cite{witte_high-entropy_2019, dokala_switching_2026, cedervall_phase_2021} Deviations in the ZFC and FC curves for LaM7O$_3$ are observed above $T_\mathrm{N}$, at about 160 K. Since no secondary phase is observed in the PXRD data for this sample, this additional irreversibility is attributed to some degree of magnetic inhomogeneity.\cite{Pramanik_spin_2024} The broad hump in the FC susceptibility below the ordering temperature is relatively clear for LaM7O$_3$, which hosts a non-magnetic A-site, meaning that the susceptibility curves describe the behaviour of the B-site lattice. However, as the number of 4f electrons is increased (La < La$_{\sfrac{1}{2}}$Gd$_{\sfrac{1}{2}}$ < A5 $\approx$ Gd), the low temperature susceptibility is dominated by the paramagnetic A-site species and features describing the B-site moments are not obviously discernable. Temperature dependent magnetisation measurements were also performed in larger magnetic fields (\textit{H} = 1 kOe) to verify the Curie-Weiss behaviour of the systems (Figure S4). At high temperatures, above the $T_\mathrm{N}$, the inverse susceptibility for all compounds is linear, suggesting that the compounds are paramagnetic in that temperature region with no short-range correlations persisting. The Curie-Weiss temperatures (Table \ref{mag_values}) range from $-$9(2) K for GdM7O$_3$ to +14(2) K for A5M7O$_3$, varying in both magnitude and sign.


In addition, isothermal magnetisation measurements were carried out at 2 K, below their magnetic ordering temperatures (Figure \ref{fgr:magnetisation}). All compounds exhibit non-linear magnetisation behaviour and increase continuously with the applied field and without clear saturation at the largest measured fields of 50 kOe. The magnetisation increases across the series as the A-site valence electrons increase, from $M_\mathrm{50\,kOe, \, La}=$ 3.7(1) emu/g, $M_\mathrm{50\,kOe, \, La\frac{1}{2}Gd\frac{1}{2}}=$ 54.5(1) emu/g, $M_\mathrm{50\,kOe, \, A5}=$ 68.5(1) emu/g and $M_\mathrm{50\,kOe, \, Gd}=$ 114.2(1) emu/g. Moreover, an exchange bias-type behaviour is observed in the FC data for GdM7O$_3$ (Figure S5), although further analysis is not in the scope of this present work.

\section*{Discussion}

Here, we have demonstrated the synthesis of a single perovskite phase hosting a total of twelve cationic species over the A- and B- sites, in addition to three other compositionally complex compounds. Despite the numerous elements present, our results show that the compounds have compositional homogeneity considering both their average structures and on microscopic scale. Multicomponent solid solutions have previously been reported with high configurational entropy, which is generally agreed to be $\geq$1.61\textit{R}.\cite{aamlid_understanding_2023} Commonly, these compounds adopt the perovskite structure-type, although series of pyrochlore, spinel and rocksalt type structures have also been explored.\cite{johnstone_entropy_2022, mao_new_2020, jiang_probing_2021, zhang_long-range_2019} The compounds reported here all have orthorhombic symmetry, likely due to the larger differences in the average A- and B-site radii. This is compensated by symmetry-lowering octahedral tilts, from the archetypical cubic system, which is also observed in the simple parent rare-earth transition metal perovskites.

The lower symmetry \textit{Pnma} space group permits additional degrees of freedom, particularly of the A-site cation in the \textit{x} and \textit{z} directions, allowing the A-site to shift to more favourable positions (Figure \ref{fgr:intro}b). This is evident in this series of compounds in which the La$^{3+}$ cation has an atomic position significantly closer to the mirror plane, (0.5234(2), 0.25, -0.0078(3)), compared to Gd$^{3+}$ (0.4364(2), 0.25, 0.0154(3)). An implicit consequence of incorporating smaller A-site cations is the rotation of the \ce{[BO6]} octahedra. Although the oxygen atomic coordinates could be more accurately determined by, for example, neutron diffraction studies, broad trends can be discussed from our X-ray diffraction results. LaM7O$_3$ demonstrates the least distorted angles: $\angle$B$-$O$-$B = 168.5(1)\textdegree\ and 162.3(7)\textdegree. However, although Gd$^{3+}$ has the smallest radius and displays more tilted octahedra ($\angle$B$-$O$-$B = 151.9(6) \textdegree\ and 149.6(5)\textdegree), it is not a linear relationship between the A-site radius and the $\angle$B$-$O$-$B tilt. Other factors, including first and second Jahn-Teller distortions, are known to influence tilt angles.\cite{lufaso_jahnteller_2004, alonso_evolution_2000} However, the ambiguity in the B-cation valancies of our compounds precludes further speculation about this.

\begin{figure}[t]
\centering
  \includegraphics{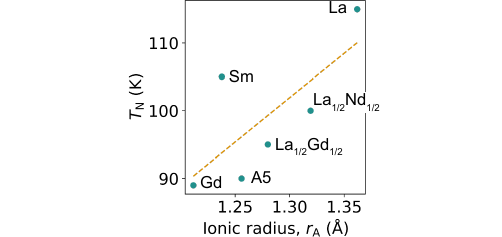}
  \caption{The magnetic ordering temperature ($T_\mathrm{N}$) as a function of average A-site ionic radius ($r_\mathrm{A}$). Values for SmM7O$_3$ and La$_{\sfrac{1}{2}}$Nd$_{\sfrac{1}{2}}$M7O$_3$ obtained from literature.\cite{dokala_switching_2026, cedervall_phase_2021} The dashed line is a guide to the eye.}
  \label{fgr:ra_tn_comp}
\end{figure}

Following the structural changes as a function of the A-site average radius is also beneficial to understand the magnetic behaviour of these compounds. From the magnetometry results, all our compounds magnetically order between 115 and 89 K, correlating well with a decrease in the A-site radius (Figure \ref{fgr:ra_tn_comp}), both for our compounds, and for previously reported compounds with AM7O$_3$ stoichiometry.\cite{dokala_switching_2026, cedervall_phase_2021} This trend is in agreement with both reported simple\cite{lei_general_2013, bertaut_etude_1966} and complex perovskite systems.\cite{witte_high-entropy_2019, sarkar_high_2023, das_comprehensive_2023} However, the range of ordering temperatures (26 K) is narrower compared to systems with a single, or two ordered, cations on the B-site, which can vary by over 100 K.\cite{booth_investigation_2009, kimura_distorted_2003} By shifting to smaller A-site radii, the \ce{[BO6]} octahedra compensate by rotating, resulting in B$-$O$-$B angles which deviate further from 180\textdegree. In turn, the exchange pathways are potentially reduced in efficiency, for example, through less complete orbital overlap and decreasing covalency of the B$-$O bonds.\cite{asai_55_2005} 



A substantial increase in the effective magnetic moment (µ$_\mathrm{eff}$) is observed with decreasing $r_\mathrm{A}$, from 3.1(1) µ$_B$, where A = La, to 8.9(1) µ$_B$, for A = Gd. Considering the A- and B-sites independently, the effective moment arising from the B-site is challenging to determine due to the number species and ambiguity in oxidation states for some of the metals. Even considering the spin-only magnetic moment, which exclusively considers the number of unpaired spins, would still require knowing the exact valence of each species. We have previously attempted to use X-ray photoelectron spectroscopy (XPS) to determine the oxidation states of similar compositionally complex oxide compounds, however, we found the results to be ambiguous, with some cations likely adopting mixed valence states.\cite{clulow_phase_2024} If +3 oxidation state is considered for all B-site species except Ti and Cu, the spin-only moment is approximately 3-4 µ$_\mathrm{B}$, similar to µ$_\mathrm{eff}$ obtained for LaM7O$_3$, which has magnetic cations solely on the B-site. The effective moment of the A-site is easier to predict as it contains trivalent 4f rare earth species. Assuming a similar B-site effective moment for all our compounds, i.e. the value determined experimentally for LaM7O$_3$, the experimentally measured total effective moments are in line with the expected values, $\sqrt{(\mu_\mathrm{eff, \; A}^2+\mu_\mathrm{eff, \; B}^2)}$ (Table \ref{mag_values}). 


Introducing a compositionally complex A-site still allows for long-range magnetic order, despite the large number of potential competing interactions present in this system. This has been observed where solely A- or B-site compositional disorder has been introduced, but also in a handful of other multi-site disordered bulk compounds\cite{witte_high-entropy_2019} and thin films.\cite{Bhattacharya_mean_2026} Contrary to potential presuppositions, it is significantly more common to report magnetic ordering in these compositionally complex oxides, with only a few compounds reporting spin-glass behaviour from magnetometry data,\cite{clulow_phase_2024} or diffuse scattering from neutron diffraction measurements.\cite{arndt_magnetic_2024} The size disorder parameter ($\sigma ^2$), for A5M7O$_3$ is slightly lower than for La$_{\sfrac{1}{2}}$Gd$_{\sfrac{1}{2}}$M7O$_3$ (Table \ref{tbl:UC_param}), suggesting that within the temperature regimes for which the A-site sublattice is paramagnetic, there may be less variation in the structural distortions. This is also evident by regarding La$_{\sfrac{1}{2}}$Nd$_{\sfrac{1}{2}}$M7O$_3$ which has smaller size disorder ($\sigma ^2 =$ 0.0020) than all of our multi-A-site materials and a higher magnetic ordering temperature ($T_\mathrm{N} =$ 100 K, $r_\mathrm{La\frac{1}{2}Nd\frac{1}{2}} =$ 1.32 Å),\cite{cedervall_phase_2021} although it is lower than SmM7O$_3$ which has a smaller $r_\mathrm{A}$ ($T_\mathrm{N} =$ 105 K, $r_\mathrm{Sm} =$ 1.24 Å).\cite{dokala_switching_2026}
Given the antagonistic functionality between decreasing $r_\mathrm{A}$ and $\sigma ^2$,\cite{das_comprehensive_2023} it is conceivable to design and maintain fine control over the ordering temperatures of materials by utilising compositionally complex compounds to select both the average A-site radius and the collective variations in the radii.  


\section*{Conclusions}
We have presented the synthesis and structural characterisation of four compositionally complex compounds, adopting single phase perovskite-type structures from PXRD, including a previously unreported compound with twelve cations across the A- and B-sites. EDS and SEM data suggest that the compounds are homogenous on a micro-length scale. Bulk magnetometry data reveals all the compounds magnetically order as ferrimagnets, with the A-site unpaired electrons, average A-site radius, and A-site size disorder all influencing the characteristics of the materials.

This work demonstrates the benefits of incorporating several cations into a single system to tailor magnetic behaviour. 
In particular, it motivates further investigations into using a multi-cationic A-site to influence the magnetic transition temperatures. Through intentional synthesis, we can consider not only the valency and number of A-site cations incorporated, but also the average and spread of radii of the constituent components.  

\section*{Author contributions}
M. G. synthesised the samples and carried out the powder X-ray diffraction experiments and analyses. M. G. and R. C. carried out the energy dispersive X-ray measurements and scanning electron microscopy images. R. K. D. and R. M. carried out and analysed the magnetic measurements. M. G. and R. C. wrote the paper with contributions from all the authors.


\section*{Conflicts of interest}
There are no conflicts to declare.

\section*{Data availability}


The data that support the findings of this article are not publicly available. the data are available on request from the authors upon reasonable request.

\section*{Acknowledgements}

R.C and M.G acknowledge financial support from the Olle Engkvist Stiftelse (Grant No. 229-0390), the Åforsk foundation (Grant No. 24-349) and the Magnus Bergvall foundation (Grant No. 2025-638). R.K.D and R.M. thank the Olle Engkvist Stiftelse (Grant No. 224-0046) for financial support. The authors acknowledge Myfab Uppsala for providing facilities and experimental support. Myfab is funded by the Swedish Research Council (2019-00207) as a national research infrastructure.



\balance


\bibliography{Mag_HEPs} 

@article{glazer_classification_1972,
	title = {The classification of tilted octahedra in perovskites},
	volume = {28},
	issn = {05677408},
	url = {https://scripts.iucr.org/cgi-bin/paper?S0567740872007976},
	doi = {10.1107/S0567740872007976},
	number = {11},
	urldate = {2023-10-31},
	journal = {Acta Crystallographica Section B Structural Crystallography and Crystal Chemistry},
	author = {Glazer, A. M.},
	month = nov,
	year = {1972},
	pages = {3384--3392},
}

@article{Goldschmidt1926,
	title = {Die {Gesetze} der {Krystallochemie}},
	volume = {14},
	journal = {Naturwissenschaften},
	author = {Goldschmidt, V. M.},
	year = {1926},
	pages = {477--485},
}

@article{clulow_phase_2024,
	title = {Phase {Stability} and {Magnetic} {Properties} of {Compositionally} {Complex} \textit{n} = 2 {Ruddlesden}–{Popper} {Perovskites}},
	volume = {63},
	copyright = {https://creativecommons.org/licenses/by/4.0/},
	issn = {0020-1669, 1520-510X},
	url = {https://pubs.acs.org/doi/10.1021/acs.inorgchem.3c04277},
	doi = {10.1021/acs.inorgchem.3c04277},
	language = {en},
	number = {15},
	urldate = {2024-09-06},
	journal = {Inorganic Chemistry},
	author = {Clulow, Rebecca and Pramanik, Prativa and Stolpe, Amanda and Joshi, Deep C. and Mathieu, Roland and Henry, Paul F. and Sahlberg, Martin},
	month = apr,
	year = {2024},
	pages = {6616--6625},
}

@article{pramanik_spin_2024,
	title = {Spin glass states in multicomponent layered perovskites},
	volume = {14},
	issn = {2045-2322},
	url = {https://www.nature.com/articles/s41598-024-53896-2},
	doi = {10.1038/s41598-024-53896-2},
	language = {en},
	number = {1},
	urldate = {2024-09-06},
	journal = {Scientific Reports},
	author = {Pramanik, P. and Clulow, R. and Joshi, D. C. and Stolpe, A. and Berastegui, P. and Sahlberg, M. and Mathieu, R.},
	month = feb,
	year = {2024},
	pages = {3382},
}

@article{cedervall_phase_2021,
	title = {Phase stability and structural transitions in compositionally complex {LnMO3} perovskites},
	volume = {300},
	issn = {00224596},
	url = {https://linkinghub.elsevier.com/retrieve/pii/S0022459621002589},
	doi = {10.1016/j.jssc.2021.122213},
	language = {en},
	urldate = {2024-09-06},
	journal = {Journal of Solid State Chemistry},
	author = {Cedervall, Johan and Clulow, Rebecca and Boström, Hanna L.B. and Joshi, Deep C. and Andersson, Mikael S. and Mathieu, Roland and Beran, Premysl and Smith, Ronald I. and Tseng, Jo-Chi and Sahlberg, Martin and Berastegui, Pedro and Shafeie, Samrand},
	month = aug,
	year = {2021},
	pages = {122213},
}

@article{coelho_topas_2018,
	title = {\textit{{TOPAS}} and \textit{{TOPAS}-{Academic}}: an optimization program integrating computer algebra and crystallographic objects written in {C}++},
	volume = {51},
	issn = {1600-5767},
	shorttitle = {\textit{{TOPAS}} and \textit{{TOPAS}-{Academic}}},
	url = {https://journals.iucr.org/paper?S1600576718000183},
	doi = {10.1107/S1600576718000183},
	number = {1},
	urldate = {2025-12-15},
	journal = {Journal of Applied Crystallography},
	author = {Coelho, Alan A.},
	month = feb,
	year = {2018},
	pages = {210--218},
}

@article{goto_ferroelectricity_2004,
	title = {Ferroelectricity and {Giant} {Magnetocapacitance} in {Perovskite} {Rare}-{Earth} {Manganites}},
	volume = {92},
	copyright = {http://link.aps.org/licenses/aps-default-license},
	issn = {0031-9007, 1079-7114},
	url = {https://link.aps.org/doi/10.1103/PhysRevLett.92.257201},
	doi = {10.1103/PhysRevLett.92.257201},
	language = {en},
	number = {25},
	urldate = {2026-05-05},
	journal = {Physical Review Letters},
	author = {Goto, T. and Kimura, T. and Lawes, G. and Ramirez, A. P. and Tokura, Y.},
	month = jun,
	year = {2004},
	pages = {257201},
}

@article{kimura_magnetoelectric_2005,
	title = {Magnetoelectric phase diagrams of orthorhombic {R} {MnO} 3 ( {R} = {Gd} , {Tb}, and {Dy})},
	volume = {71},
	copyright = {http://link.aps.org/licenses/aps-default-license},
	issn = {1098-0121, 1550-235X},
	url = {https://link.aps.org/doi/10.1103/PhysRevB.71.224425},
	doi = {10.1103/PhysRevB.71.224425},
	language = {en},
	number = {22},
	urldate = {2026-05-05},
	journal = {Physical Review B},
	author = {Kimura, T. and Lawes, G. and Goto, T. and Tokura, Y. and Ramirez, A. P.},
	month = jun,
	year = {2005},
	pages = {224425},
}

@article{witte_high-entropy_2019,
	title = {High-entropy oxides: {An} emerging prospect for magnetic rare-earth transition metal perovskites},
	volume = {3},
	issn = {2475-9953},
	shorttitle = {High-entropy oxides},
	url = {https://link.aps.org/doi/10.1103/PhysRevMaterials.3.034406},
	doi = {10.1103/PhysRevMaterials.3.034406},
	language = {en},
	number = {3},
	urldate = {2026-05-05},
	journal = {Physical Review Materials},
	author = {Witte, Ralf and Sarkar, Abhishek and Kruk, Robert and Eggert, Benedikt and Brand, Richard A. and Wende, Heiko and Hahn, Horst},
	month = mar,
	year = {2019},
	pages = {034406},
}

@article{zhang_magnetic_2022,
	title = {Magnetic properties and giant cryogenic magnetocaloric effect in {B}-site ordered antiferromagnetic {Gd2MgTiO6} double perovskite oxide},
	volume = {226},
	issn = {13596454},
	url = {https://linkinghub.elsevier.com/retrieve/pii/S1359645422000544},
	doi = {10.1016/j.actamat.2022.117669},
	language = {en},
	urldate = {2026-05-05},
	journal = {Acta Materialia},
	author = {Zhang, Yikun and Tian, Yun and Zhang, Zhenqian and Jia, Youshun and Zhang, Bin and Jiang, Minqiang and Wang, Jiang and Ren, Zhongming},
	month = mar,
	year = {2022},
	pages = {117669},
}

@article{khosrozadeh_complex_2024,
	title = {Complex impedance spectroscopy, dielectric response, and magnetic properties of the {La0}.7 {Sr0}.{3BO3} ({B} = {Mn}, {Fe}, {Co}, or {Ni}) perovskite oxides},
	volume = {50},
	issn = {02728842},
	url = {https://linkinghub.elsevier.com/retrieve/pii/S0272884223031486},
	doi = {10.1016/j.ceramint.2023.10.105},
	language = {en},
	number = {1},
	urldate = {2026-05-05},
	journal = {Ceramics International},
	author = {Khosrozadeh, M. and Mabhouti, Kh. and Norouzzadeh, P. and Naderali, R.},
	month = jan,
	year = {2024},
	pages = {315--328},
}

@article{belguenoune_structural_2026,
	title = {Structural {Characteristics} and {Optical} {Properties} of {La1}−{xYxFeO3} {Perovskites} for {Optoelectronic} and {Nonlinear} {Optical} {Applications}},
	issn = {1574-1443, 1574-1451},
	url = {https://link.springer.com/10.1007/s10904-026-04303-y},
	doi = {10.1007/s10904-026-04303-y},
	language = {en},
	urldate = {2026-05-05},
	journal = {Journal of Inorganic and Organometallic Polymers and Materials},
	author = {Belguenoune, Ahmed and Ouldhamadouche, Nadir and Bassaid, Salah and Dehbi, Abdelkader and Dammak, Sameh and Ben Rhaiem, Abdallah},
	month = apr,
	year = {2026},
}

@article{rajendran_tri-doped_2020,
	title = {Tri-{Doped} {BaCeO}$_{\textrm{3}}$ –{BaZrO}$_{\textrm{3}}$ as a {Chemically} {Stable} {Electrolyte} with {High} {Proton}-{Conductivity} for {Intermediate} {Temperature} {Solid} {Oxide} {Electrolysis} {Cells} ({SOECs})},
	volume = {12},
	copyright = {https://doi.org/10.15223/policy-029},
	issn = {1944-8244, 1944-8252},
	url = {https://pubs.acs.org/doi/10.1021/acsami.0c12532},
	doi = {10.1021/acsami.0c12532},
	language = {en},
	number = {34},
	urldate = {2026-05-07},
	journal = {ACS Applied Materials \& Interfaces},
	author = {Rajendran, Sathish and Thangavel, Naresh Kumar and Ding, Hanping and Ding, Yi and Ding, Dong and Reddy Arava, Leela Mohana},
	month = aug,
	year = {2020},
	pages = {38275--38284},
}

@article{hong_microstructural_2019,
	title = {Microstructural evolution and mechanical properties of ({Mg},{Co},{Ni},{Cu},{Zn}){O} high‐entropy ceramics},
	volume = {102},
	issn = {0002-7820, 1551-2916},
	url = {https://ceramics.onlinelibrary.wiley.com/doi/10.1111/jace.16075},
	doi = {10.1111/jace.16075},
	language = {en},
	number = {4},
	urldate = {2026-05-07},
	journal = {Journal of the American Ceramic Society},
	author = {Hong, Weichen and Chen, Fei and Shen, Qiang and Han, Young‐Hwan and Fahrenholtz, William G. and Zhang, Lianmeng},
	month = apr,
	year = {2019},
	pages = {2228--2237},
}

@article{zhao_effect_2013,
	title = {Effect of chemical and hydrostatic pressures on structural and magnetic properties of rare-earth orthoferrites: a first-principles study},
	volume = {25},
	copyright = {http://iopscience.iop.org/info/page/text-and-data-mining},
	issn = {0953-8984, 1361-648X},
	shorttitle = {Effect of chemical and hydrostatic pressures on structural and magnetic properties of rare-earth orthoferrites},
	url = {https://iopscience.iop.org/article/10.1088/0953-8984/25/46/466002},
	doi = {10.1088/0953-8984/25/46/466002},
	number = {46},
	urldate = {2026-05-07},
	journal = {Journal of Physics: Condensed Matter},
	author = {Zhao, Hong Jian and Ren, Wei and Yang, Yurong and Chen, Xiang Ming and Bellaiche, L},
	month = nov,
	year = {2013},
	pages = {466002},
}

@article{zhou_intrinsic_2010,
	title = {Intrinsic structural distortion and superexchange interaction in the orthorhombic rare-earth perovskites {R} {CrO} 3},
	volume = {81},
	copyright = {http://link.aps.org/licenses/aps-default-license},
	issn = {1098-0121, 1550-235X},
	url = {https://link.aps.org/doi/10.1103/PhysRevB.81.214115},
	doi = {10.1103/PhysRevB.81.214115},
	language = {en},
	number = {21},
	urldate = {2026-05-13},
	journal = {Physical Review B},
	author = {Zhou, J.-S. and Alonso, J. A. and Pomjakushin, V. and Goodenough, J. B. and Ren, Y. and Yan, J.-Q. and Cheng, J.-G.},
	month = jun,
	year = {2010},
	pages = {214115},
}

@article{terakura_magnetism_2007,
	title = {Magnetism, orbital ordering and lattice distortion in perovskite transition-metal oxides},
	volume = {52},
	copyright = {https://www.elsevier.com/tdm/userlicense/1.0/},
	issn = {00796425},
	url = {https://linkinghub.elsevier.com/retrieve/pii/S0079642506000727},
	doi = {10.1016/j.pmatsci.2006.10.007},
	language = {en},
	number = {2-3},
	urldate = {2026-05-13},
	journal = {Progress in Materials Science},
	author = {Terakura, K},
	month = feb,
	year = {2007},
	pages = {388--400},
}

@article{lyubutin_dependence_1999,
	title = {Dependence of exchange interactions on chemical bond angle in a structural series: {Cubic} perovskite-rhombic orthoferrite-rhombohedral hematite},
	volume = {88},
	copyright = {https://www.springernature.com/gp/researchers/text-and-data-mining},
	issn = {1063-7761, 1090-6509},
	shorttitle = {Dependence of exchange interactions on chemical bond angle in a structural series},
	url = {https://link.springer.com/10.1134/1.558833},
	doi = {10.1134/1.558833},
	language = {en},
	number = {3},
	urldate = {2026-05-13},
	journal = {Journal of Experimental and Theoretical Physics},
	author = {Lyubutin, I. S. and Dmitrieva, T. V. and Stepin, A. S.},
	month = mar,
	year = {1999},
	pages = {590--597},
}

@article{treves_dependence_1965,
	title = {Dependence of superexchange interaction on {Fe3}+-{O2}−-{Fe3}+ linkage angle},
	volume = {18},
	copyright = {https://www.elsevier.com/tdm/userlicense/1.0/},
	issn = {00319163},
	url = {https://linkinghub.elsevier.com/retrieve/pii/0031916365902945},
	doi = {10.1016/0031-9163(65)90294-5},
	language = {en},
	number = {3},
	urldate = {2026-05-13},
	journal = {Physics Letters},
	author = {Treves, D. and Eibschütz, M. and Coppens, P.},
	month = sep,
	year = {1965},
	pages = {216--217},
}

@article{zhou_intrinsic_2008,
	title = {Intrinsic structural distortion in orthorhombic perovskite oxides},
	volume = {77},
	copyright = {http://link.aps.org/licenses/aps-default-license},
	issn = {1098-0121, 1550-235X},
	url = {https://link.aps.org/doi/10.1103/PhysRevB.77.132104},
	doi = {10.1103/PhysRevB.77.132104},
	language = {en},
	number = {13},
	urldate = {2026-05-13},
	journal = {Physical Review B},
	author = {Zhou, J.-S. and Goodenough, J. B.},
	month = apr,
	year = {2008},
	pages = {132104},
}

@article{das_comprehensive_2023,
	title = {Comprehensive analysis on the effect of ionic size and size disorder parameter in high entropy stabilized ferromagnetic manganite perovskite},
	volume = {7},
	issn = {2475-9953},
	url = {https://link.aps.org/doi/10.1103/PhysRevMaterials.7.024411},
	doi = {10.1103/PhysRevMaterials.7.024411},
	language = {en},
	number = {2},
	urldate = {2026-05-13},
	journal = {Physical Review Materials},
	author = {Das, Radhamadhab and Pal, Sudip and Bhattacharya, Sudipa and Chowdhury, Shreyashi and K. K., Supin and Vasundhara, M. and Gayen, Arup and Seikh, Md. Motin},
	month = feb,
	year = {2023},
	pages = {024411},
}

@article{das_comparative_2021,
	title = {A comparative magnetic behaviour of conventional and high entropy double perovskites: {La2MnCoO6} and ({La0}.{4Y0}.{4Ca0}.{4Sr0}.{4Ba0}.4){MnCoO6}},
	volume = {538},
	issn = {03048853},
	shorttitle = {A comparative magnetic behaviour of conventional and high entropy double perovskites},
	url = {https://linkinghub.elsevier.com/retrieve/pii/S0304885321005436},
	doi = {10.1016/j.jmmm.2021.168267},
	language = {en},
	urldate = {2026-05-18},
	journal = {Journal of Magnetism and Magnetic Materials},
	author = {Das, Radhamadhab and Bhattacharya, Sudipa and Haque, Ariful and Ghosh, Debamalya and Lebedev, Oleg I. and Gayen, Arup and Seikh, Md. Motin},
	month = nov,
	year = {2021},
	pages = {168267},
}

@article{witte_magnetic_2020,
	title = {Magnetic properties of rare-earth and transition metal based perovskite type high entropy oxides},
	volume = {127},
	issn = {0021-8979, 1089-7550},
	url = {https://pubs.aip.org/jap/article/127/18/185109/280823/Magnetic-properties-of-rare-earth-and-transition},
	doi = {10.1063/5.0004125},
	language = {en},
	number = {18},
	urldate = {2026-05-18},
	journal = {Journal of Applied Physics},
	author = {Witte, Ralf and Sarkar, Abhishek and Velasco, Leonardo and Kruk, Robert and Brand, Richard A. and Eggert, Benedikt and Ollefs, Katharina and Weschke, Eugen and Wende, Heiko and Hahn, Horst},
	month = may,
	year = {2020},
	pages = {185109},
}

@article{rodriguez-martinez_structural_1999,
	title = {Structural {Effects} of {Cation} {Size} {Variance} in {Magnetoresistive} {Manganese} {Oxide} {Perovskites}},
	volume = {11},
	issn = {0897-4756, 1520-5002},
	url = {https://pubs.acs.org/doi/10.1021/cm980759y},
	doi = {10.1021/cm980759y},
	language = {en},
	number = {6},
	urldate = {2026-05-20},
	journal = {Chemistry of Materials},
	author = {Rodríguez-Martínez, Lide M. and Attfield, J. Paul},
	month = jun,
	year = {1999},
	pages = {1504--1509},
}

@article{maignan_size_1999,
	title = {Size mismatch: {A} crucial factor for generating a spin-glass insulator in manganites},
	volume = {60},
	copyright = {http://link.aps.org/licenses/aps-default-license},
	issn = {0163-1829, 1095-3795},
	shorttitle = {Size mismatch},
	url = {https://link.aps.org/doi/10.1103/PhysRevB.60.15214},
	doi = {10.1103/PhysRevB.60.15214},
	language = {en},
	number = {22},
	urldate = {2026-05-20},
	journal = {Physical Review B},
	author = {Maignan, A. and Martin, C. and Van Tendeloo, G. and Hervieu, M. and Raveau, B.},
	month = dec,
	year = {1999},
	pages = {15214--15219},
}

@article{aamlid_understanding_2023,
	title = {Understanding the {Role} of {Entropy} in {High} {Entropy} {Oxides}},
	volume = {145},
	copyright = {https://doi.org/10.15223/policy-029},
	issn = {0002-7863, 1520-5126},
	url = {https://pubs.acs.org/doi/10.1021/jacs.2c11608},
	doi = {10.1021/jacs.2c11608},
	language = {en},
	number = {11},
	urldate = {2026-06-03},
	journal = {Journal of the American Chemical Society},
	author = {Aamlid, Solveig S. and Oudah, Mohamed and Rottler, Jörg and Hallas, Alannah M.},
	month = mar,
	year = {2023},
	pages = {5991--6006},
}

@article{johnstone_entropy_2022,
	title = {Entropy {Engineering} and {Tunable} {Magnetic} {Order} in the {Spinel} {High}-{Entropy} {Oxide}},
	volume = {144},
	copyright = {https://doi.org/10.15223/policy-029},
	issn = {0002-7863, 1520-5126},
	url = {https://pubs.acs.org/doi/10.1021/jacs.2c06768},
	doi = {10.1021/jacs.2c06768},
	language = {en},
	number = {45},
	urldate = {2026-06-03},
	journal = {Journal of the American Chemical Society},
	author = {Johnstone, Graham H. J. and González-Rivas, Mario U. and Taddei, Keith M. and Sutarto, Ronny and Sawatzky, George A. and Green, Robert J. and Oudah, Mohamed and Hallas, Alannah M.},
	month = nov,
	year = {2022},
	pages = {20590--20600},
}

@article{mao_new_2020,
	title = {A new class of spinel high-entropy oxides with controllable magnetic properties},
	volume = {497},
	issn = {03048853},
	url = {https://linkinghub.elsevier.com/retrieve/pii/S0304885319325740},
	doi = {10.1016/j.jmmm.2019.165884},
	language = {en},
	urldate = {2026-06-03},
	journal = {Journal of Magnetism and Magnetic Materials},
	author = {Mao, Aiqin and Xiang, Hou-Zheng and Zhang, Zhan-Guo and Kuramoto, Koji and Zhang, Hui and Jia, Yanggang},
	month = mar,
	year = {2020},
	pages = {165884},
}

@article{jiang_probing_2021,
	title = {Probing the {Local} {Site} {Disorder} and {Distortion} in {Pyrochlore} {High}-{Entropy} {Oxides}},
	volume = {143},
	copyright = {https://doi.org/10.15223/policy-029},
	issn = {0002-7863, 1520-5126},
	url = {https://pubs.acs.org/doi/10.1021/jacs.0c10739},
	doi = {10.1021/jacs.0c10739},
	language = {en},
	number = {11},
	urldate = {2026-06-03},
	journal = {Journal of the American Chemical Society},
	author = {Jiang, Bo and Bridges, Craig A. and Unocic, Raymond R. and Pitike, Krishna Chaitanya and Cooper, Valentino R. and Zhang, Yuanpeng and Lin, De-Ye and Page, Katharine},
	month = mar,
	year = {2021},
	pages = {4193--4204},
}

@article{zhang_long-range_2019,
	title = {Long-{Range} {Antiferromagnetic} {Order} in a {Rocksalt} {High} {Entropy} {Oxide}},
	volume = {31},
	copyright = {https://doi.org/10.15223/policy-029},
	issn = {0897-4756, 1520-5002},
	url = {https://pubs.acs.org/doi/10.1021/acs.chemmater.9b00624},
	doi = {10.1021/acs.chemmater.9b00624},
	language = {en},
	number = {10},
	urldate = {2026-06-03},
	journal = {Chemistry of Materials},
	author = {Zhang, Junjie and Yan, Jiaqiang and Calder, Stuart and Zheng, Qiang and McGuire, Michael A. and Abernathy, Douglas L. and Ren, Yang and Lapidus, Saul H. and Page, Katharine and Zheng, Hong and Freeland, John W. and Budai, John D. and Hermann, Raphael P.},
	month = may,
	year = {2019},
	pages = {3705--3711},
}

@article{sarkar_high_2023,
	title = {High {Entropy} {Approach} to {Engineer} {Strongly} {Correlated} {Functionalities} in {Manganites}},
	volume = {35},
	issn = {0935-9648, 1521-4095},
	url = {https://advanced.onlinelibrary.wiley.com/doi/10.1002/adma.202207436},
	doi = {10.1002/adma.202207436},
	language = {en},
	number = {2},
	urldate = {2026-06-05},
	journal = {Advanced Materials},
	author = {Sarkar, Abhishek and Wang, Di and Kante, Mohana V. and Eiselt, Luis and Trouillet, Vanessa and Iankevich, Gleb and Zhao, Zhibo and Bhattacharya, Subramshu S. and Hahn, Horst and Kruk, Robert},
	month = jan,
	year = {2023},
	pages = {2207436},
}

@article{lei_general_2013,
	title = {General synthesis of rare-earth orthochromites with quasi-hollow nanostructures and their magnetic properties},
	volume = {1},
	issn = {2050-7488, 2050-7496},
	url = {https://xlink.rsc.org/?DOI=c3ta12281f},
	doi = {10.1039/c3ta12281f},
	language = {en},
	number = {38},
	urldate = {2026-06-05},
	journal = {Journal of Materials Chemistry A},
	author = {Lei, Shuijin and Liu, Lei and Wang, Chunying and Wang, Chuanning and Guo, Donghai and Zeng, Suyuan and Cheng, Baochang and Xiao, Yanhe and Zhou, Lang},
	year = {2013},
	pages = {11982},
}

@article{bertaut_etude_1966,
	title = {Etude des propri\&\#233;t\&\#233;s magn\&\#233;tostatiques et des structures magn\&\#233;tiques des chromites des terres rares et d'yttrium},
	volume = {2},
	copyright = {https://ieeexplore.ieee.org/Xplorehelp/downloads/license-information/IEEE.html},
	issn = {0018-9464, 1941-0069},
	url = {https://ieeexplore.ieee.org/document/1065951/},
	doi = {10.1109/TMAG.1966.1065951},
	number = {3},
	urldate = {2026-06-05},
	journal = {IEEE Transactions on Magnetics},
	author = {Bertaut, E. and Mareschal, J. and De Vries, G. and Aleonard, R. and Pauthenet, R. and Rebouillat, J. and Zarubicka, V.},
	month = sep,
	year = {1966},
	pages = {453--458},
}

@article{arndt_magnetic_2024,
	title = {Magnetic structure and properties of the compositionally complex perovskite ({Y}$_{\textrm{0.2}}$ {La}$_{\textrm{0.2}}$ {Pr}$_{\textrm{0.2}}$ {Nd}$_{\textrm{0.2}}$ {Tb}$_{\textrm{0.2}}$ ){MnO}$_{\textrm{3}}$},
	volume = {12},
	issn = {2050-7526, 2050-7534},
	url = {https://xlink.rsc.org/?DOI=D4TC01411A},
	doi = {10.1039/D4TC01411A},
	language = {en},
	number = {34},
	urldate = {2026-06-05},
	journal = {Journal of Materials Chemistry C},
	author = {Arndt, Nathan D. and Musicó, Brianna L. and Parui, Kausturi and Sahebkar, Keon and Zhang, Qiang and Mazza, Alessandro R. and Butala, Megan M. and Keppens, Veerle and Ward, T. Zac and Need, Ryan F.},
	year = {2024},
	pages = {13474--13484},
}

@article{booth_investigation_2009,
	title = {An investigation of structural, magnetic and dielectric properties of {R2NiMnO6} ({R}=rare earth, {Y})},
	volume = {44},
	copyright = {https://www.elsevier.com/tdm/userlicense/1.0/},
	issn = {00255408},
	url = {https://linkinghub.elsevier.com/retrieve/pii/S0025540809000609},
	doi = {10.1016/j.materresbull.2009.02.003},
	language = {en},
	number = {7},
	urldate = {2026-06-09},
	journal = {Materials Research Bulletin},
	author = {Booth, R.J. and Fillman, R. and Whitaker, H. and Nag, Abanti and Tiwari, R.M. and Ramanujachary, K.V. and Gopalakrishnan, J. and Lofland, S.E.},
	month = jul,
	year = {2009},
	pages = {1559--1564},
}

@article{kimura_distorted_2003,
	title = {Distorted perovskite with e g 1 configuration as a frustrated spin system},
	volume = {68},
	copyright = {http://link.aps.org/licenses/aps-default-license},
	issn = {0163-1829, 1095-3795},
	url = {https://link.aps.org/doi/10.1103/PhysRevB.68.060403},
	doi = {10.1103/PhysRevB.68.060403},
	language = {en},
	number = {6},
	urldate = {2026-06-09},
	journal = {Physical Review B},
	author = {Kimura, T. and Ishihara, S. and Shintani, H. and Arima, T. and Takahashi, K. T. and Ishizaka, K. and Tokura, Y.},
	month = aug,
	year = {2003},
	pages = {060403},
}

@article{asai_55_2005,
	title = {$^{\textrm{55}}$ {Mn} {NMR} in {Ferromagnetic} {Perovskites} {RENi}$_{\textrm{0.5}}$ {Mn}$_{\textrm{0.5}}$ {O}$_{\textrm{3}}$ ({RE} = {Rare} {Earth} {Element})},
	volume = {74},
	issn = {0031-9015, 1347-4073},
	url = {https://journals.jps.jp/doi/10.1143/JPSJ.74.1289},
	doi = {10.1143/JPSJ.74.1289},
	language = {en},
	number = {4},
	urldate = {2026-06-09},
	journal = {Journal of the Physical Society of Japan},
	author = {Asai, Kichizo and Kobayashi, Norifumi and Bairo, Tomoaki and Kaneko, Naoki and Kobayashi, Yoshihiko and Suzuki, Masaru and Satoh, Yasumasa and Mizoguchi, Moriji},
	month = apr,
	year = {2005},
	pages = {1289--1296},
}

@article{bhattacharya_mean_2026,
	title = {Mean field magnetism and spin frustration in a double perovskite oxide with compositional complexity},
	volume = {7},
	issn = {2662-4443},
	url = {https://www.nature.com/articles/s43246-026-01135-8},
	doi = {10.1038/s43246-026-01135-8},
	language = {en},
	number = {1},
	urldate = {2026-06-10},
	journal = {Communications Materials},
	author = {Bhattacharya, Nandana and Dokala, Ravi Kiran and Chowdhury, Sourav and Joshi, Suresh Chandra and Dey, Subha and Dey, Jayjit Kumar and Nandy, Subhajit and Pérez-Salinas, Daniel and Valvidares, Manuel and Hoesch, Moritz and Mathieu, Roland and Middey, Srimanta},
	month = mar,
	year = {2026},
	pages = {130},
}

@article{baloch_extending_2021,
	title = {Extending {Shannon}'s ionic radii database using machine learning},
	volume = {5},
	issn = {2475-9953},
	url = {https://link.aps.org/doi/10.1103/PhysRevMaterials.5.043804},
	doi = {10.1103/PhysRevMaterials.5.043804},
	language = {en},
	number = {4},
	urldate = {2026-06-24},
	journal = {Physical Review Materials},
	author = {Baloch, Ahmer A. B. and Alqahtani, Saad M. and Mumtaz, Faisal and Muqaibel, Ali H. and Rashkeev, Sergey N. and Alharbi, Fahhad H.},
	month = apr,
	year = {2021},
	pages = {043804},
}

@article{lufaso_jahnteller_2004,
	title = {Jahn–{Teller} distortions, cation ordering and octahedral tilting in perovskites},
	volume = {60},
	issn = {0108-7681},
	url = {https://journals.iucr.org/paper?S0108768103026661},
	doi = {10.1107/S0108768103026661},
	number = {1},
	urldate = {2026-06-25},
	journal = {Acta Crystallographica Section B Structural Science},
	author = {Lufaso, Michael W. and Woodward, Patrick M.},
	month = feb,
	year = {2004},
	pages = {10--20},
}

@article{alonso_evolution_2000,
	title = {Evolution of the {Jahn}−{Teller} {Distortion} of {MnO}$_{\textrm{6}}$ {Octahedra} in {RMnO}$_{\textrm{3}}$ {Perovskites} ({R} = {Pr}, {Nd}, {Dy}, {Tb}, {Ho}, {Er}, {Y}): {A} {Neutron} {Diffraction} {Study}},
	volume = {39},
	issn = {0020-1669, 1520-510X},
	shorttitle = {Evolution of the {Jahn}−{Teller} {Distortion} of {MnO}$_{\textrm{6}}$ {Octahedra} in {RMnO}$_{\textrm{3}}$ {Perovskites} ({R} = {Pr}, {Nd}, {Dy}, {Tb}, {Ho}, {Er}, {Y})},
	url = {https://pubs.acs.org/doi/10.1021/ic990921e},
	doi = {10.1021/ic990921e},
	language = {en},
	number = {5},
	urldate = {2026-06-25},
	journal = {Inorganic Chemistry},
	author = {Alonso, J. A. and Martínez-Lope, M. J. and Casais, M. T. and Fernández-Díaz, M. T.},
	month = mar,
	year = {2000},
	pages = {917--923},
}

@article{dokala_switching_2026,
	title = {Switching magnetic spin states using small magnetic fields in compositionally complex {Sm} ( {Ti} , {Cr} , {Mn} , {Fe} , {Co} , {Ni} , {Cu} ) {O} 3},
	volume = {114},
	issn = {2469-9950, 2469-9969},
	url = {https://link.aps.org/doi/10.1103/b77s-6l6d},
	doi = {10.1103/b77s-6l6d},
	language = {en},
	number = {3},
	urldate = {2026-08-05},
	journal = {Physical Review B},
	author = {Dokala, R. K. and Geers, M. and Nordblad, P. and Clulow, R. and Mathieu, R.},
	month = jul,
	year = {2026},
	pages = {034404},
}

@article{tomioka_global_2004,
	title = {Global phase diagram of perovskite manganites in the plane of quenched disorder versus one-electron bandwidth},
	volume = {70},
	copyright = {http://link.aps.org/licenses/aps-default-license},
	issn = {1098-0121, 1550-235X},
	url = {https://link.aps.org/doi/10.1103/PhysRevB.70.014432},
	doi = {10.1103/PhysRevB.70.014432},
	language = {en},
	number = {1},
	urldate = {2026-09-01},
	journal = {Physical Review B},
	author = {Tomioka, Y. and Tokura, Y.},
	month = jul,
	year = {2004},
	pages = {014432},
}
\bibliographystyle{rsc} 
\end{document}